\documentclass[a4paper,english,a4paper, amsfonts, amssymb, amsmath, reprint, showkeys, nofootinbib, twoside]{revtex4-1}
\usepackage[utf8]{inputenc}
\usepackage{color}
\definecolor{document_fontcolor}{rgb}{1, 0, 0}
\color{document_fontcolor}
\usepackage{babel}
\usepackage{array}
\usepackage{multirow}
\usepackage{amsbsy}
\usepackage{amstext}
\usepackage{graphicx}
\usepackage{rotating}
\usepackage{esint}
\usepackage[unicode=true,
 bookmarks=false,
 breaklinks=false,pdfborder={0 0 1},backref=section,colorlinks=false]
 {hyperref}
\hypersetup{
 pdftex}

\makeatletter

\pdfpageheight\paperheight
\pdfpagewidth\paperwidth

\providecommand{\tabularnewline}{\\}

\usepackage{siunitx}
\usepackage{chemformula}
\usepackage{array}

\makeatother

\begin{document}
\title{Interplay between exchange split Dirac and Rashba-type surface states
in MnBi$_{2}$Te$_{4}$/BiTeI interface}
\author{N.L. Zaitsev}
\affiliation{Institute of Molecule and Crystal Physics, Ufa Research Center of
Russian Academy of Sciences, 450075, Ufa, Russia}
\author{I. P. Rusinov}
\affiliation{Tomsk State University, 634050 Tomsk, Russia}
\author{T. V. Menshchikova}
\affiliation{Tomsk State University, 634050 Tomsk, Russia}
\author{E. V. Chulkov}
\affiliation{Saint Petersburg State University, 198504 Saint Petersburg, Russia}
\affiliation{Departamento de Polímeros y Materiales Avanzados: Física, Química
y Tecnología, Facultad de Ciencias Químicas, Universidad del País
Vasco UPV/EHU, 20080 Donostia-San Sebastián, Basque Country, Spain}
\affiliation{Donostia International Physics Center (DIPC), 20018 Donostia-San Sebastián,
Basque Country, Spain}
\affiliation{Tomsk State University, 634050 Tomsk, Russia}
\begin{abstract}
Based on the \emph{ab initio} calculations, we study the electronic
structure of the \ch{BiTeI}/\ch{MnBi2Te4} heterostructure interface
composed of the anti-ferromagnetic topological insulator \ch{MnBi2Te4}
and the polar semiconductor trilayer BiTeI. We found significant difference
in electronic properties at different types of contact between substrate
and the overlayer. While the case of Te-Te interface forms natural
expansion of the substrate, when Dirac cone state locates mostly in
the polar overlayer region and undergoes slight exchange splitting,
Te-I contact is the source of four-band state contributed by the substrate
Dirac cone and Rashba-type state of the polar trilayer. Owing to magnetic
proximity, the pair of Kramers degeneracies for this state are lifted,
what produces Hall response in transport regime. We believe, our findings
provide new opportunities to construct novel type spintronic devices.
\end{abstract}
\maketitle

\section{INTRODUCTION}

Interplay between spin-orbit interaction and magnetism attracts a
lot of attention owing to the impact on band topology and electron
transport phenomena~\citep{Jungwirth2016,Basov2017,MTI}. In the
case of asymmetric bulk and surface systems, the spin-orbit effects
produce Rashba spin splitting of bulk and surface bands~\citep{Bychkov1984,Bihlmayer_2015},
what is exploited in proposed spin-field transistor~\citep{Datta-Das,spintronics}.
Another example is quantum spin Hall effect reflected in the formation
of Dirac cone states with ``spin filtering'' transport property
on the boundaries of topological insulators (TIs)~\citep{Hasan-Kane_RMP}.
Introduction of magnetism enriches the complexity and noteworthiness
of the systems with mentioned spin-orbital phenomena via the breaking
time-reversal symmetry and thus lifting degeneracies of Rashba states
and Dirac cones. It forms additional topological band gap what is
the source of spin-based transport phenomena like recently proposed
chiral orbital magnetization effect~\citep{Lux2018}, what allows
applying such systems in spintronic devices~\citep{spintronics,RevModPhys.91.035004,AHE}
and quantum computation~\citep{Majorana,Quant_comput}.

One possible strategy of further research of interrelation between
magnetic and spin-orbit effects is based on the design of complex
heterostructures with both these contributions. Owing to the weak
chemical interaction between building blocks, the van der Waals compounds
provide a suitable platform to design the systems with desired\textcolor{red}{{}
}properties~\citep{Geim2013,Novoselov_2016} via the employing molecular
beam epitaxy or mechanical exfoliation techniques~\citep{VDW_Fabric}.

The ideal ingredients for design of complex heterostructures with
both magnetic and spin-orbital effects are antiferromagnetic topological
insulator \ch{MnBi2Te4} and polar semiconductor BiTeI. The former
is magnetic semiconductor \citep{Otrokov2019_Nat} composed of seven
layer (7L) blocks coupled by Van der Waals forces along the {[}0001{]}
direction.\textcolor{red}{{} }This magnetic topological insulator has
been proposed as efficient platform for magnetic spintronics~\citep{Otrokov_2017,PhysRevLett.122.107202},
containing exchange split bands on the (0001) cleavage plane and providing
effect of magnetic proximity. Another constituent, the polar semiconductor
BiTeI, is built up by polar trilayers, and is characterized by giant
Rashba-type spin splitting of both bulk gap edge states and surface
state~\citep{Ishizaka2011,PhysRevLett.108.246802}. Excellent matching
of in-plane crystal cell parameters for both constituents prevents
dislocations or Moiré pattern effects during formation of the interface.

Here, we report a density functional theory (DFT) study of the van
der Waals heterostructure composed by antiferromagnetic topological
insulator \ch{MnBi2Te4}(0001) surface (MBT) and polar semiconductor
BiTeI trilayer resulting in formation of BiTeI/MBT interface. The
Te--Te contact case forms non-magnetic extension of \ch{MnBi2Te4}
pristine surface by BiTeI trilayer, what is expressed in spatial shifting
of exchange splitted Dirac cone surface state into the overlayer region.
This behavior is caused by strong spin-orbit contribution both in
substrate and overlayer, and by the absence of strong perturbation
of electrostatic potential over BiTeI trilayer deposition. Also it
is accompanied by the shrinking of the exchange band gap of the Dirac
cone and its downward shifting at overlayer deposition.

In the case of Te--I contact, the energy spectrum near the Fermi
level is formed by four-band state composed by Rashba-type state of
the overlayer and Dirac state of the magnetic substrate surface. In
the vicinity of the Brillouin zone center, there are two types of
features: 1) hybridized band gap, owing to interaction of these two
states, and 2) two exchange gaps, owing to magnetic proximity with
the substrate. The latter features are the source of intrinsic Hall
conductivity, due to time-reversal symmetry breaking, what allows
applying this state in spintronic devices. This finding demonstrates
another way to form hybridization gap between Rashba-type state and
Dirac cone state, which also previously observed in pristine MBT~\citep{Liang2022}
and \ch{MnBi6Te10}~\citep{PhysRevB.102.245136} surfaces.

\begin{figure*}[t]
\includegraphics[height=0.6\textheight]{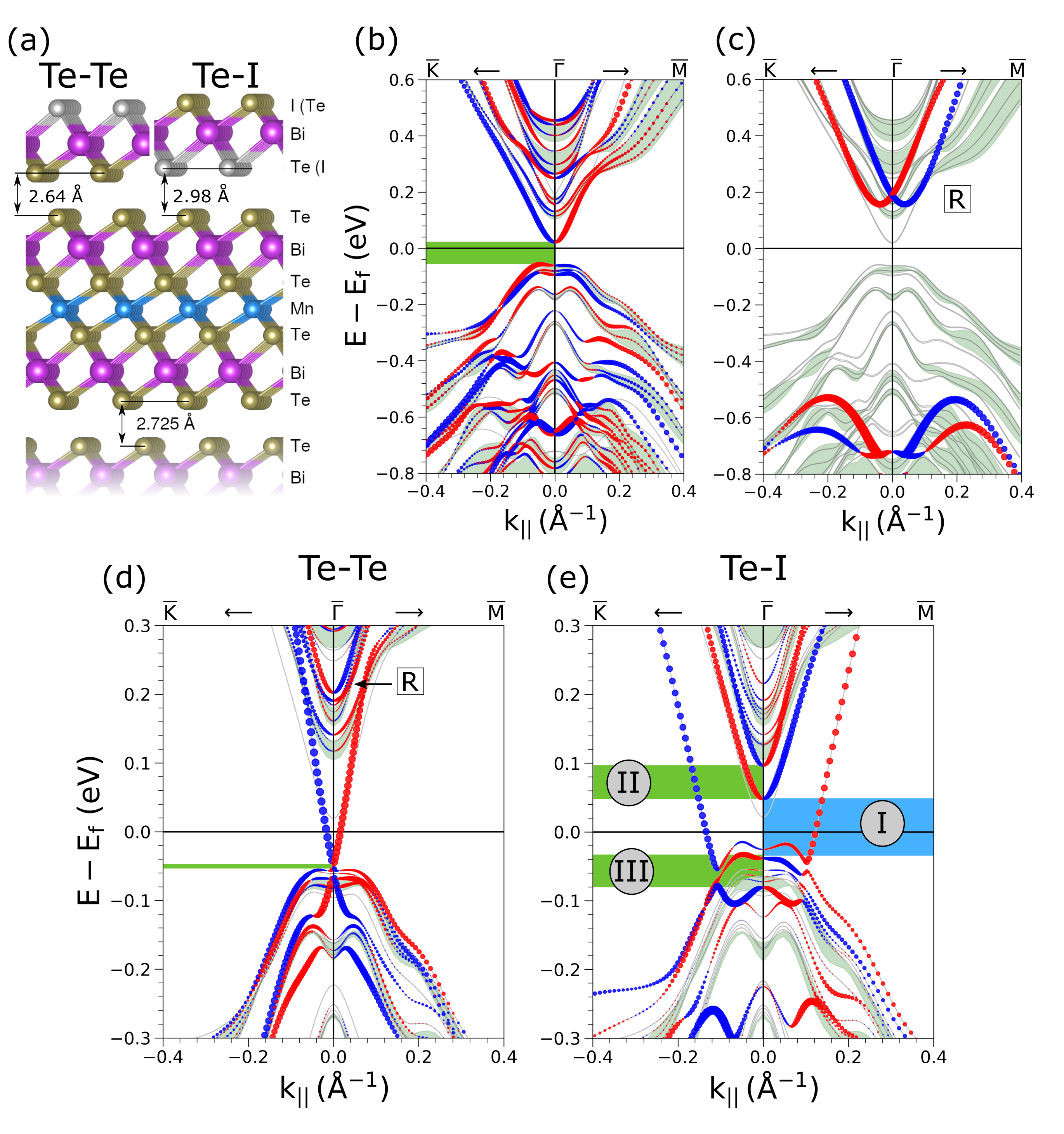}

\caption{\label{fig:parts}(a) Shematic geometrical structure of the BiTeI/MBT
interface with different orientations of BiTeI trilayer. \emph{In-plane}
spin-resolved band structure calculated with large separation of 12 \AA \  
between (b) pristine $\mathrm{MnBi_{2}Te_{4}(0001)}$ surface and
(c) BiTeI trilayer. Spin-resolved electron spectrum for (d) Te-Te
and (e) Te-I interfaces in the case of equilibrium structures. The
value and direction of \emph{in-plane} spin projection are coded by
circle size and color (red and blue). The  gray area represents bulk
projected bands, and the black rectangular emphasizes the trivial
surface states of $\mathrm{MnBi_{2}Te_{4}(0001)}$. For panels (b),
(d) and (e), the magnetic echange and hybridization energy gap at
the center of SBZ are denoted by green (blue) rectangles. In panel
(e), the hybridization energy gap is denoted as I, and two exhange
gaps~--- as II and III. In panels (c) and (d), the unoccupied Rashba
state are denoted by R symbol.}
\end{figure*}

\section{Calculation details}

The crystal structure of $\mathrm{MnBi_{2}Te_{4}}$ is characterized
by lattice parameters $a=\SI{4.33}{\angstrom}$ and $c=\SI{40.93}{\angstrom}$
\citep{lee_crystal_2013}. The same \emph{a }parameter is used for
BiTeI trilayer placed on top of MBT (Fig. \ref{fig:parts}a). The
semi-infinite MBT was represented by periodically repeated slabs of
6 septuple layers in width, with vacuum imposed in \emph{z} direction
along surface normal. The parallel and antiparallel aligning of the
\ch{MnBi2Te4}(0001) surface normal with the BiTeI trilayer dipole
moment (directed towards the Te layer) was considered. Hereinafter,
we denote these two cases as Te--I and Te--Te interfaces (see Fig.
\ref{fig:parts} a). We considered type of junction between the substrate
and the overlayer which is similar to one of between adjacent seven-layer
blocks inside the MBT substrate.

The equilibrium vertical separation $d_{0}$ between MBT and trilayer
was determined from relaxation of interlayer distances of the first
MBT septuple layer along with BiTeI, whereas the rest of the slab
was fixed in bulk geometry. The structural optimization was performed
within the PBE-D3 scheme \citep{grimme_consistent_2010,grimme_effect_2011},
which incorporates an empirical correction to include dispersion forces
on top of the PBE functional using the projector augmented-wave (PAW)
method~\citep{paw1,paw2} implemented in VASP~\citep{VASP}. The
Hamiltonian contained scalar relativistic corrections, and the spin-orbit
coupling (SOC) was taken into account by the second variation method~\citep{SOC}.
Note, that influence of spin-orbit coupling on force field is quite
noticeable. The elongation of the equilibrium vertical separation
$d_{0}$ due to SOC reaches \SI{0.1}{\angstrom} in case of the Te-I
interface providing $d_0=\SI{2.98}{\angstrom}$, while it is half
as much for the Te-Te case with $d_0=\SI{2.64}{\angstrom}$.

\textit{Ab initio} electronic structure calculations were performed
within the DFT as implemented in the OpenMX (version 3.8) code \citep{_openmx_38}.
The linear combination of localized pseudoatomic orbitals \citep{ozaki_variationally_2003,ozaki_numerical_2004,ozaki_efficient_2005}
was used to construct the basis functions. The norm-conserving pseudopotential
\citep{troullier_efficient_1991} was taken as a replacement for deep
core potential. The generalized gradient approximation Perdew-Burke-Ernzerhof
(PBE) functional \citep{perdew_generalized-gradient_1996} was applied
for the exchange-correlation energy. The basis functions were set
as follows: Mn6.0-\emph{s3p3d2}, Te(I)7.0-\emph{s3p3d2f1} and Bi8.0-\emph{s3p3d2f1},
namely, 3 primitive orbitals for each \emph{s} and \emph{p} channels
and 2 primitive orbital for the \emph{d} channel with the cutoff radius
of 6.0~a.u. were used to define Mn atoms etc. The real-space grid
for numerical integration and solution of the Poisson equation was
set to 200~Ry of the cutoff energy. The total-energy convergence
criterion was $3\cdot10^{-5}$~eV. The surface Brillouin zone (SBZ)
of the supercell was sampled with a $7\times7$ mesh of \textbf{k}
points.

The calculated in-plane spin-resolved band structure of BiTeI/MBT
interface with 12\,\AA \  separation between the MBT surface and
the BiTeI trilayer (to eliminate any interactions between them) is
shown on Fig. \ref{fig:parts}~b,~c. The panels depict the bands
localized within the surface septuple layer of MBT (Fig. \ref{fig:parts}~b)
and within the BiTeI trilayer (Fig. \ref{fig:parts}~c), respectively.
In contrast to the nonmagnetic case of the structurally similar $\mathrm{PbBi_{2}Te_{4}}$
and $\mathrm{PbSb_{2}Te_{4}}$ surfaces, where Kramers degeneracies
are located within the projected band gap \citep{eremeev_new_2015,Eremeev2012_natcomm},
the presence of a magnetic exchange field of MBT lifts this degeneracy
in topological surface states (TSS) what forms the gap of $\sim80\,\mathrm{meV}$
(Fig. \ref{fig:parts}~b). Thus, the upper part of the Dirac cone
lies within the projected band gap above the Fermi level, whereas
the lower part with a flattened vertex resides just above the valence
band \citep{Otrokov2019_Nat,shikin_sample-dependent_2021}. Note,
that at energy of $\sim-0.65\,\mathrm{eV}$, there are trivial surface
states (SS) of MBT (Fig.~\ref{fig:parts}~b) which are heavily involved
into interaction with BiTeI, as will be seen further. In turn, the
degenerate point of highly split Rashba state of BiTeI trilayer lies
above the Fermi level and moreover overlaps with the bulk projected
bands of MBT (Fig.~\ref{fig:parts}~c).

\begin{figure*}
\includegraphics[height=0.5\textheight]{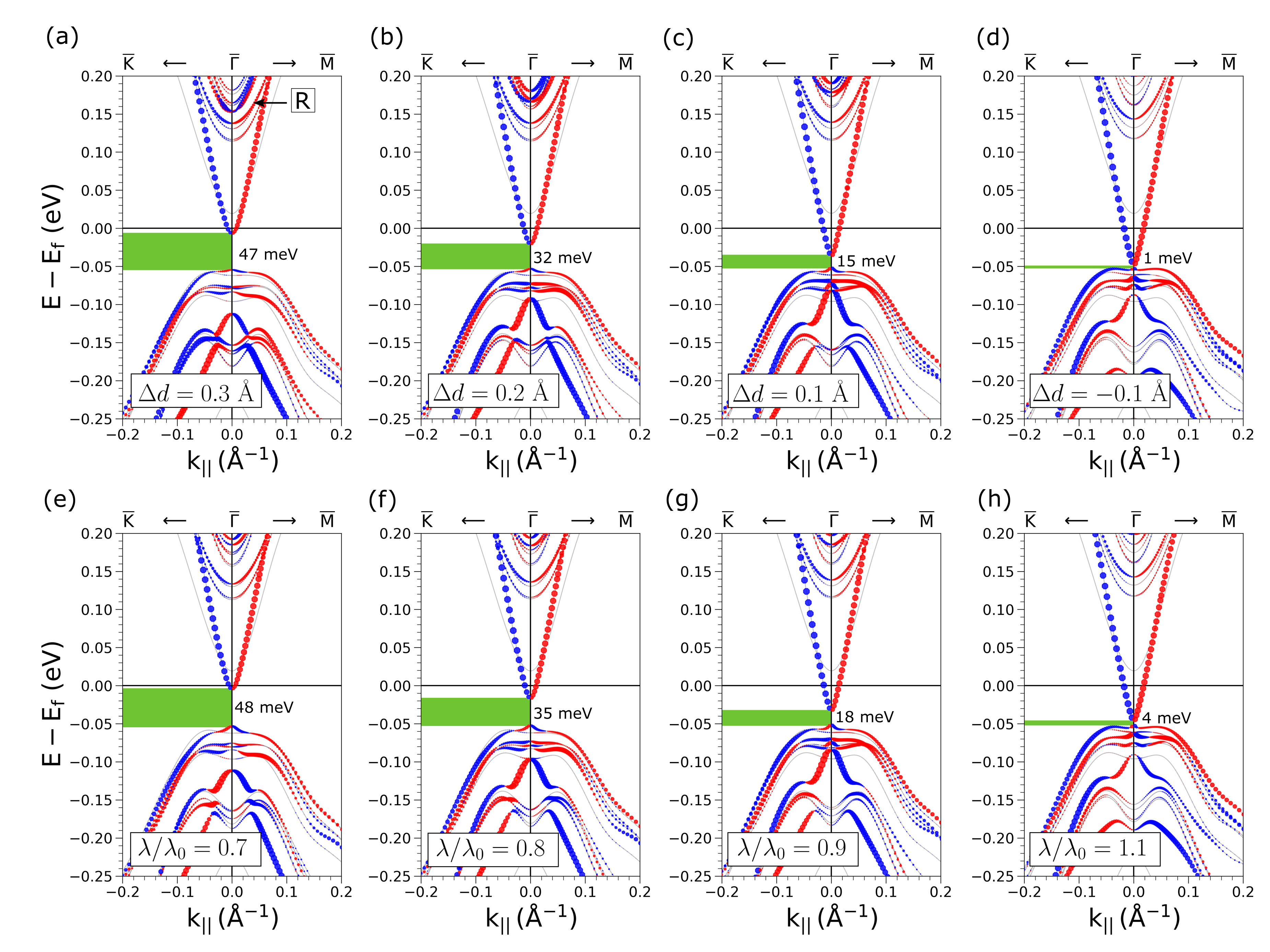}

\caption{\label{fig:Te-side}Surface spin-resolved electronic structure of
Te--Te interface (a--d) for different separations between the MBT
surface and trilayer with respect to equilibrium geometry, $\Delta d=d-d_{0}$,
and (e--h) for different spin-orbit coupling strength of Bi and Te
\emph{p}-states of BiTeI trilayer. The natural SOC contribution gives
$\lambda/\lambda_{0}=1$. On the panels, magnetic exchange gap at
SBZ center are reflected by green areas. For each panel, the gap width
is denoted from the right side of the green area. In panel (a), the
subsurface Rashba state is denoted by R symbol.}
\end{figure*}

\section{Results}

The relaxed Te-Te and Te-I interfaces have similar values of total
energy, where the first one is 0.1~eV more favorable than the second
one. Also, the first interface is characterized by the $\sim0.3\,\textrm{Å}$
shorter interlayer distance between trilayer and MBT (Te-Te interface
spacing~--- $d_{0}=2.64\,\mathrm{\text{Å}}$, Te-I~--- $d_{0}=2.98\,\mathrm{\text{Å}}$)
and is a bit shorter than vdW spacing of MBT subtrate ($d_{vdW}=2.725\,\mathrm{\text{Å}}$)
{[}see also Fig.~\ref{fig:parts}~a{]}. The other inter-plane distances
are tolerant to the interface type.

\emph{Ab-initio} spin-resolved surface electronic spectrum of Te-Te
interface (Fig.~\ref{fig:parts}~d) has notable changes with respect
to the pristine MBT surface (Fig.~\ref{fig:parts}~b) and nonmagnetic
$\mathrm{BiTeI/PbSb_{2}Te_{4}}$ interface with the same layer stacking
\citep{eremeev_new_2015}. Namely, the tiny exchange Dirac gap (see
green rectangle on Fig.~\ref{fig:parts}~d) of a few meV width takes
place, locates just above the valence band maximum. The shifted down
Dirac state has increased velocity. With that, at the SBZ center the
lower part of the Dirac cone overlaps with the set of weak surface
states inherited from the highest MBT bulk valence state, due to elctrostatic
field near the surface. In turn, the Rashba-type state resides at
energy of $\sim0.2\,\textrm{eV}$ lessening their momentum splitting.
Note, that BiTeI overlayer on MBT prone to pull the surface states
into itself, like in the case of adsorption on Au(111) \citep{zaitsev_spinorbit_2019}
or $\mathrm{PbSb_{2}Te_{4}(0001)}$ \citep{eremeev_new_2015} surfaces.

The another trends can be seen in electronic structure of Te-I interface
(see Fig.~\ref{fig:parts}~e). In the area near the Fermi level
the set of spin-polarized states appears. As will be discussed afterwards
in detail, they are separated by energy gaps of different nature.
The two gaps (II and III) {[}green color rectangles{]} are of exchange
type and are originated from the presence of magnetic MBT substrate.
The highest (lowest) exchange gap is of $\sim49$~meV (of $\sim42$~meV)
width. Another type gap (I) is crossed by the Fermi level and is of
$\sim86$~meV width (blue color rectangle). It has hybridization
character and is induced by interaction of the surface cone state
of MBT and the Rashba-type state of the overlayer. The gap of same
nature has been observed previously at consideration of non-magnetic
$\mathrm{BiTeI/PbSb_{2}Te_{4}}$ surface heterostructure \citep{eremeev_new_2015}.
In such a way, Te-I interface forms single four-band state composed
of two-band Dirac and Rashba-type states. Out of the area of the SBZ
center, where hybridization gap is formed, the discussed four-band
state dispersion is inherited from these spin-orbit contributions,
separately. It should be noted, that due to presence of the valence
band maximum which plays a role of charge reservoir, the potential
gradient near the surface region produces the additional surface states
which are involved into the interaction with four-band state under
consideration, which can be regarded as simplification of low-energy
surface electronic spectrum for this type of interface.

\subsection*{Te--Te interface}

\begin{figure}
\includegraphics[height=0.3\textheight]{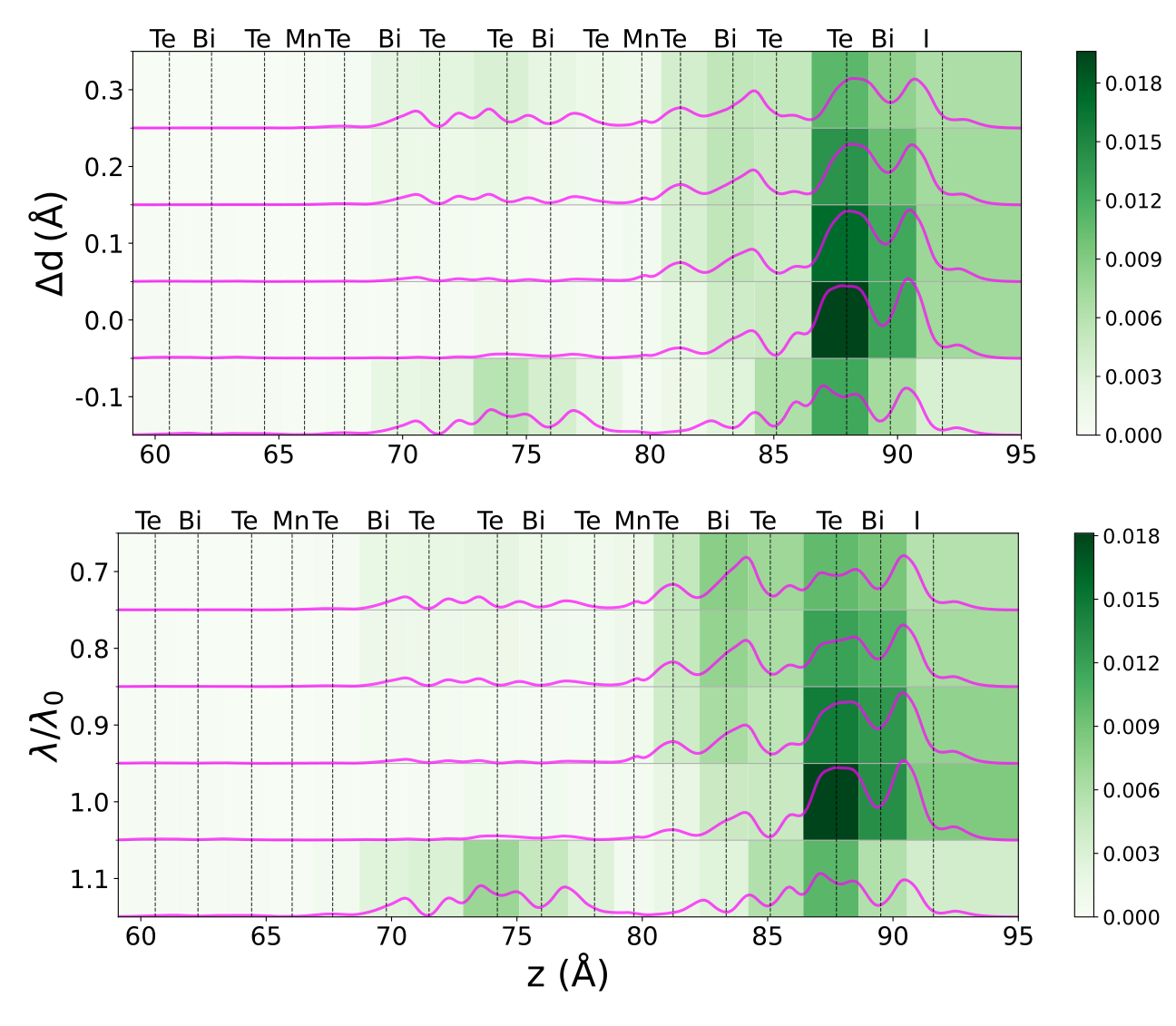}

\caption{\label{fig:chgtete} Te-Te interface: charge density distribution
of the upper part of the Dirac surface state (purple curves) at the
$\overline{\Gamma}$-point as function of out-of-plane direction (integrated
over $xy$-plane) for different vertical spacing between the MBT surface
and trilayer (top panel) and spin-orbit coupling strength, $\lambda/\lambda_{0}$
(bottom panel) {[}see also Fig. \ref{fig:Te-side}{]}. The integral
of charge density inside the vicinity of adjoined atomic layers are
color-coded by intensities of green.}

\end{figure}

The origins of the calculated electronic structure of the equilibrium
Te-Te interface can be clarified by varying the vertical separation,
$d$, between MBT and polar trilayer what changes interaction between
these building blocks. Also, electronic spectra at increased $d$
correspond to the cases of additional intercalated atoms into the
van der Waals region \citep{Eremeev_2012_exp,Eremeev_2018_exp,Eremeev_2021_exp}.
At $d$=$4\,\textrm{Å}$, the interaction between substrate and overlayer
is weak, and the bands alignment has no substantial changes with respsect
to fully decoupled case presented on the Fig.~\ref{fig:parts}~c.
However, this weak interaction induces notable $\sim0.3\,\textrm{eV}$
upward shift of Rashba-type surface states. The decreasing distance
$d$ reduces the Rashba-type splitting of bands and enforces the mentioned
energy shift accompanied by charge density relocation into the upper
septuple block of the MBT substrate. When the spacing increases and
becomes to be of $0.3\,\textrm{Å}$ greater than the equilibrium
one ($\Delta d=d-d_{0}=0.3\,\mathrm{\text{Å}}$), the former Rashba
state dimishes surface character and completely declines their momentum
splitting (see Fig.~\ref{fig:Te-side} a). Herewith, in the area
of Fermi level the exchange gap of the Dirac states reduces owing
to down energy shift of the upper part of the cone, and the apex of
lower part lies at energy of $-0.11\,\textrm{\text{eV}}$ at the $\overline{\Gamma}$-point.
At further reduction of $d$, the formation of almost gapless Dirac
state is revealed which is hybridized with the bulk bands near the
SBZ center forming set of surface resonances (Fig. \ref{fig:Te-side}
c). The minimal gap is obsered at equilibrium distance, $d=d_{0}$,
(Fig.~\ref{fig:parts}~d). The subsequent shrinking of the interlayer
distance, $d$, shifts the topological surface states down in the
energy scale (Fig. \ref{fig:Te-side}d), albeit the exchange gap size
is still negligible. Note, that the TSS wave function tends to be
localized in the trilayer when approaching to MBT similar to the case
of BiTeI/Au(111) interface \citep{zaitsev_spinorbit_2019}. However,
under further reducing $d$ the TSS pulls back to the MBT surface
(Fig.~\ref{fig:chgtete}) due to the proximity of the TSS and the
highest valence bulk states continuum what produces the resonant character
of TSS in the energy area near the exchange gap.

The distance dependent surface electronic structure evolves in a similar
way as it happens at artificial SOC modulation on Bi and Te atoms
of the trilayer (see Fig.~\ref{fig:Te-side}). At increasing interlayer
distance by $\Delta d=0.1\,\textrm{Å}$, surface states shift in
a similar fashion as at reducing the SOC factor by 10\%, $\lambda/\lambda_{0}=0.9$
(Fig.~\ref{fig:Te-side}~c,g). Note, that it is enough to modulate
SOC factor of \emph{p}-states only, since the TSS are predominantly
composed by this type of orbitals. Again, under compression, $\Delta d=-0.1\,\textrm{Å}$
, (Fig.~\ref{fig:Te-side}~d) or, otherwise, at increased SOC factor,
$\lambda/\lambda_{0}=1.1$ , the lower part of the cone shifts down
in the energy scale (Fig. \ref{fig:Te-side}~h). Moreover, the wave
function of the upper part of cone is maximally localized on the trilayer
at equilibrium distance and natural SOC contribution, $\lambda/\lambda_{0}=1$.
With increasing $\lambda/\lambda_{0}$ , the Dirac cone charge localization
is extruded back to the MBT in the same way as at approaching BiTeI
trilayer close to MBT surafce (Fig.~\ref{fig:chgtete}).

The effect described just above is originated by antiparallel directed
trilayer dipole moment which induces a potential gradient on the MBT
surface. The same has been previously observed under additional surface
doping \citep{shikin_sample-dependent_2021} when the surface negative
charge reduces the exchange gap in pristine MBT \citep{shikin_sample-dependent_2021},
due to downshift of the upper part of the split Dirac cone. Such relationship
of spin-orbit interaction and electric field effects has been revealed
in various materials \citep{chen_electric-field_2018,lin_interface-based_2019,maryenko_interplay_2021}.

For this type termination, the formation of the exchange gap in the
Dirac state is the source of the range of transport phenomena: half-quantized
Hall conductivity\citep{QHZ-2008}, anomalous Hall effect\citep{Mogi_2019,Deng_2020}
and topological magnetoelectric effect\citep{Wang_2015,Zirnstein-Rosenow}.

\subsection*{Te--I interface}

The inverted polarity of the trilayer in the Te-I interface leads
to the parallel orientation of the dipole moment of this building
block with respect to the MBT surface normal. It provides the positive
potential gradient near the vacuum region. Hence, at approaching of
the trilayer closer to the MBT surface the Dirac state reallocates
inside the trilayer. Such an effect of TSS redistribution towards
the vacuum boundary has been previously observed in the case of MBT
surfaces under the positive external electric field \citep{shikin_sample-dependent_2021}.
Also, in the case of the Te-I interface over the $\Delta d$ decreasing
the Rashba-type state moves down in the spectrum. In such a way, at
$\Delta d=1\,\textrm{Å}$, this state overlaps with the lower part
of the Dirac cone immediately below the Fermi level with the formation
of hybridization between them. It manifests the single four-band composite
state persisted at subsequent decreasing of $\Delta d$.

\begin{figure*}
\includegraphics[height=0.27\textheight]{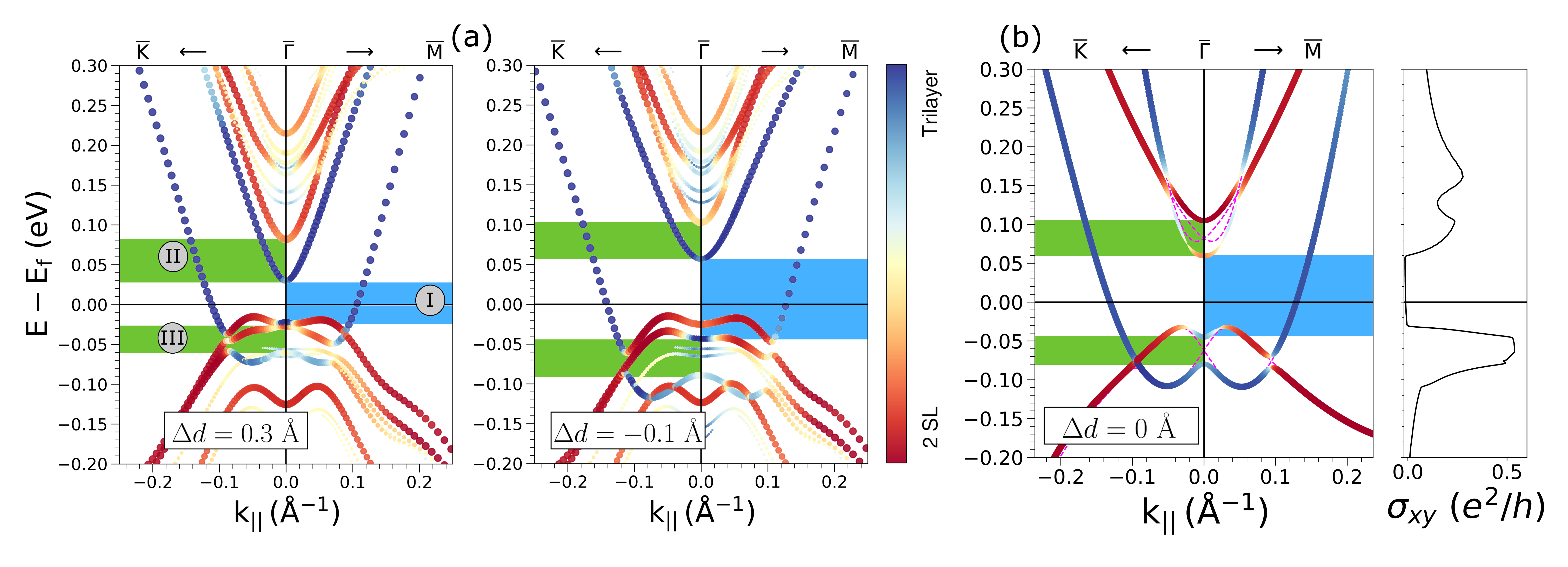}

\caption{\label{fig:I-side} (a) Surface electronic structure of the Te-I interface
for different spacings between the BiTeI overlayer and MBT surface.
The spacing is $0.3\,\mathrm{\mathsf{\textrm{Å}}}$ greater (left
panel) and $0.1\,\text{Å}$ lesser (right panel) then the equilibrium
one. The colors highlight the extent of spacial localization of the
states inside the trilayer (blue) or within the two utmost septuple
layers of MBT (red). (b) Electronic band structure (eq.~\ref{eq:Ham})
of proposed model (left panel) with parameters tabulated in Table~\ref{tab:Model_params}
(solid black lines), and without taken into account magnetic contribution,
$\hat{H_{m}}$ ($\Delta_{R}=\Delta_{D}=0$) (dashed red lines). (Right
panel) Energy dependence of Hall conductivity $\sigma_{xy}(E)=V/(2\pi)^{2}\int d^{2}\mathbf{{k}}\sigma_{xy}(\mathbf{{k}},E)$.
For all panels, the magnetic echange and hybridization energy gap
at the center of SBZ are denoted by green (blue) areas. Hybridization
energy gap is denoted as I, and two exchange gaps~--- as II and
III.}
\end{figure*}

In Fig.~\ref{fig:I-side}~a, the corresponding spectra are shown
for $\Delta d=0.3\,\textrm{Å}$ (left panel) and $\Delta d=-0.1\,\textrm{Å}$
(right panel). The approaching the trilayer closer to the MBT surface
affects the width of local band gaps of different nature. Over this
process, the hybridization gap (marked as I on the figure) is becoming
larger (from 57~meV at $\Delta d=0.3\,\textrm{Å}$, up to 100~meV
at $\Delta d=-0.1\,\textrm{Å}$), what agrees with enhancing the
interaction between building blocks. At the same time, local exchange
gaps (II, III) behave differently. The unoccupied one is shrinking
from 52~meV ($\Delta d=0.3\,\textrm{Å}$) to 45~meV ($\Delta d=-0.1\,\textrm{Å}$),
while the occupied one, on the contrary, enlarges from 34~meV to
46~meV. Such an effect says for the complexity of the interaction
between hybridization and exchange contribution for this four-band
state. Another effect of the complex hybridization of Rashba-type
and Dirac states is expressed in spectra by changing the character
of the contribution in the vicinity of the $\overline{\Gamma}$-point
for the hybridization gap edges. The unoccupied branch is composed
by the Dirac state contribution what corresponds to the localization
within the two upper SL blocks (2SL), while the occupied one is formed
by the Rashba-type contribution (localized within the trilayer block).
Out of the SBZ center, the contribution becomes inverted, i.e. unoccupied
band is contributed by the trilayer block, while the occupied one~---
by the 2SL blocks.

To address the transport properties of the composite four-band surface
state at Te-I interface, we use the simple $\mathbf{{kp}-}$model.
The model Hamiltonian (eq. \ref{eq:Ham}) is composed by two contributions.
The first one describes the linear Dirac-type states and Rashba-type
states interaction, second one~--- time-reversal symmetry breaking
magnetic contribution, $\hat{{H}}_{m}$:

\begin{equation}
\hat{H}(\mathbf{{k}})=\left(\begin{array}{cc}
\hat{{H}^{D}}(\mathbf{{k}}) & \hat{{H}}{}_{int}\\
\hat{{H}}_{int}^{\dagger} & \hat{{H}^{R}}(\mathbf{{k}})
\end{array}\right)-\hat{{H}}_{m}.\label{eq:Ham}
\end{equation}
Here, $\hat{{H}^{D}}(\mathbf{{k}})$ and $\hat{{H}^{R}}(\mathbf{{k}})$
are 2$\times$2 Hamiltonians of the Dirac and Rashba-type states localized
in the uppermost seven-layer block of the substrate and BTI overlayer,
respectively. Both the contributions have the same form, distinguishing
by parameter values:

\[
\hat{H}^{\mu}(\mathbf{{k}})=M_{0}^{\mu}+M_{1}^{\mu}k^{2}+\alpha_{\mu}(k_{x}\hat{{\sigma}}{}_{y}-k_{y}\hat{{\sigma}}{}_{x})+\frac{{\gamma_{\mu}}}{2}(k_{+}^{3}+k_{-}^{3})\hat{{\sigma}}{}_{z},
\]
where $\hat{\sigma}$ are Pauli matrices in spin space, $\mu$ identifies
the Dirac (D) or Rashba (R) part of the Hamiltonian , $M_{0}^{\mu}$
and $M_{1}^{\mu}$~are constant-energy shift and kinetic energy strength
contribution, respectively. $\alpha_{\mu}$~and $\gamma_{\mu}$ are
spin-orbit and hexagonal warping strength, respectively\citep{LFu_warp}.
Hybridization contribution, $\hat{{H}}{}_{int}$, does not depend
on momentum and takes the form $A\sigma_{0}$. Magnetic term is defined
by $\hat{{H}}_{m}=\text{{diag}}(\Delta_{D},-\Delta_{D,}\Delta_{R},-\Delta_{R}),$
where $\Delta_{\mu}$ is the strength of Zeeman contribution with
out-of-plane magnetic moment. It should be noted that the proposed
model is relevant within the small area near the $\overline{\Gamma}$-point,
where possible higher-order terms are caused by the impact of another
nearby surface and resonant states. The parameters of the Hamiltonian
have been obtained via the fitting procedure applied to \emph{ab-initio}
surface spectrum, and are shown in the Table~\ref{tab:Model_params}.
The Rashba-type part, $\hat{{H}}_{R}(\mathbf{{k}})$, has dominant
kinetic energy term, $M_{1}^{R}$, with respect to the spin-orbit
contribution strength, $\alpha_{R}$, and has opposite sign with respect
to the one of ${H}_{D}(\mathbf{{k}})$. The relative difference between
exchange parameters $\Delta_{R}$ and $\Delta_{D}$($\Delta_{D}/\Delta_{R}=3.5$)
in the Rashba-type state and linear Dirac state is in the good accordance
with aspects of localization of these states: the former is located
mostly in the trilayer block, while the latter~--- in uppermost
seven-layer block of the magnetic substrate.

On the left panel of Fig.~\ref{fig:I-side}~b, energy spectrum of
the presented model is shown for the parameters given in the Table~\ref{tab:Model_params}
( color-coded curves), and for these parameters, but without magnetic
contribution, $\Delta_{D}=\Delta_{R}=0$ (red dashed lines). The bright
feature of the presented spectra is the hybridization band gap near
the $\overline{{\Gamma}}$-point, at -0.02---0.05~eV energy range
(Fig.~\ref{fig:I-side}.a), what is formed by non-zero $A$, and
this aspect of the model exactly reproduces \emph{ab-initio} results.
As can be seen in the figure the magnetism enhances the range of features
in the spectrum. Firstly, it produces local exchange gaps in the vicinity
of the $\overline{\Gamma}$-point at $\sim$-0.08~eV and $\sim$0.1~eV
(green color on the figure), what is also in agreement with \emph{ab-initio}
results. Secondly, it is avoided crossings: at $\sim$-0.08~eV, they
are located along $\overline{{\Gamma}}-\overline{{\mathrm{{K}}}}$
direction, while along $\overline{{\Gamma}}-\overline{{\mathrm{{M}}}}$
such features are also caused by additional hexagonal warping effect.
At $\sim$0.15~eV, the formation of this type of features is artificial,
owing to limitation of $\textbf{kp}$-model producing the intersection
of Rashba-type and the linear Dirac state branches without magnetism.

\begin{table}[t]
\caption{\label{tab:Model_params}Parameters of the four-band (eq.~\ref{eq:Ham}),
obtained from the fitting {\it ab-initio} band spectrum.}

\begin{tabular}{ccc|ccc}
\multirow{1}{*}{\begin{turn}{90}
\end{turn}} & $\mu=D$ & \multirow{1}{*}{$\mu=R$} &  & $\mu=D$ & $\mu=R$\tabularnewline
\hline 
$M_{0}^{\mu}$ (eV) & 0.05 & -0.02 & $\alpha_{\mu}$(eV $\text{Å}$) & 1.42 & -1.90\tabularnewline
$M_{1}^{\mu}$ (eV$\text{\ensuremath{\mathring{A}^{2}}}$) & 2.29 & 15.66 & $\gamma_{\mu}$(eV$\text{\ensuremath{\mathring{A}^{3}}}$) & 11.85 & 26.67\tabularnewline
\cline{4-6} \cline{5-6} \cline{6-6} 
$\Delta_{\mu}$ (eV) & \multirow{1}{*}{0.035} & 0.010 & A (eV) & \multicolumn{2}{c}{0.06}\tabularnewline
\end{tabular}
\end{table}

Both time reversal symmetry breaking and surface inversion asymmetry
induce non-zero Hall conductivity. On the right panel of Fig.~\ref{fig:I-side}~b,
the energy dependence of Hall conductivity, $\sigma_{xy}$, integrated
over momentum space at each $E$. Hall conductivity has been calculated
by using anti-symmetric component of topological Berry curvature tensor
$\Omega_{n}^{xy}({k})$:

\[
\sigma_{xy}(\mathbf{{k}},E)=\frac{{e^{2}}}{h}\sum_{n}f(E_{n}-E)\Omega_{xy}^{n}(\mathbf{{k}}),
\]

where $f$ is Fermi function, and $\Omega_{n}^{xy}({\boldsymbol{k}})$
has been computed by using Kubo formula:

\[
\Omega_{xy}^{n}(\mathbf{{\mathbf{k}})=\mathrm{2}\hbar^{2}\mathrm{Im}\sum_{\mathrm{n'\ne n}}\frac{{\mathrm{\langle n|\hat{v}_{x}|n'\rangle\langle n'|\hat{v}_{y}|n\rangle}}}{\mathrm{\left[E_{n}-E{}_{n'}\right]^{2}}}},
\]

where velocity operators are $\hat{\mathrm{v}}_{x,y}=1/\hbar\,\partial\hat{H}/\partial k_{x,y}$.

The maximal intensity is located at energy where magnetic contribution
into the band dispersion is maximal, i.e. the regions of the exchange
gaps and avoided crossings. The bright peak of $\sigma_{xy}(E)$ is
located at $-0.08$~eV, decaying down to the energy scale. At energy
range of 0.08-0.18~eV, two-peak feature is shown, where the lower
peak corresponds to the exchange gap at the $\overline{{\Gamma}}$-point
and the higher~--- the discussed avoided crossing feature. In such
a way, there are two energy areas contributed to the Hall conductivity,
and they correspond to the\textcolor{red}{{} }local band gaps of lifted
Kramers degeneracies due to the magnetism.

\section{Conclusion}

In the case of Te-side MBT/BTI, the effect of the presence of the
overlayer induces spatial shift of exchange splitted Dirac cone into
the region of the BiTeI overlayer and in reducing of the exchange
gap size with respect to pristine surface of the substrate. As the
result, the surface Dirac cone mostly locates inside the polar overlayer
despite that Rashba-type spin splitting is spectral peculiarity of
BiTeI compound. Such an effect is related with bulk gap edge states
inversion of the substrate, what has been also demonstrated via the
\emph{ab-initio} calculations. In such a way, the deposition of BiTeI
trilayer on \ch{MnBi2Te4} surface can be the route to manipulation
of exchange gap size of the Dirac cone.

Between two possible side surfaces of MBT/BTI interfaces, the Te-side
interface case stands out by the formation of four-band state induced
by Rashba-type and linear Dirac cone coupling. Owing to magnetic nature
of the substrate, this state undergoes sizable exchange splitting,
what ensures intrinsic Hall conductivity contribution via the time-reversal
symmetry breaking. Herewith, the observed conductivity is not quantized
due to non-zero density of states in the corresponded energy region.
The described properties of the four-band state resemble those of
widely studied exchange-split Rashba-type state which is a useful
model to study fundamental aspects of anomalous Hall conductivity~\citep{AHE}.
Hence, one can expect the same magneto-transport phenomena for the
described four-band state of the current investigation. First of all,
additional random impurities should produce extrinsic side-jump and
skew-scattering contribution to anomalous Hall conductivity in the
I-side of the MBT/BTI interface~\citep{PhysRevB.71.224423,Onoda_PRL_2006,PhysRevLett.117.046601}.
Next, one can expect the surface anisotropic magnetoresistance effect~\citep{PhysRevB.80.134405,Wadehra2020}.
Due to strong spin-orbit coupling contribution of BiTeI overlayer,
Dzyaloshinskii-Moriya spin interaction can be produced what ensures
the formation of skyrmions and magnetic domain walls~\citep{DW},
what opens the way for manipulation of spin momentum of electron based
on recently proposed chiral orbital magnetization effect~\citep{Lux2018}.
Another type of related photo-voltaic and optical effects in the proposed
system are photo-current at zero bias voltage~\citep{PhysRevB.72.245327,APL_Pershin_2005,Ogawa2016}
or topological Kerr effect~\citep{Bang2014,PhysRevLett.117.157202}.
By organization of proximity with superconductor, it can possible
create Majorana fermion states~\citep{PhysRevLett.103.020401,PhysRevB.79.094504}
what allows one to apply the proposed heterostructure in quantum computation~\citep{Quant_comput}.
We note, the considered four-band state can be more advantage with
respect to the Rashba-type state, owing to the presence of two wide
energy ranges with contribution of Hall conductivity. As a results,
the Fermi-level can be easily pinned at this energy regions via the
surface doping.

\section{ACKNOWLEDGEMENTS}

E.V.C. acknowledges support from Saint Petersburg State University
(Grant No. ID 90383050). N.L.Z., I.P.R. and T.V.M. acknowledge support
from Russian Science Foundation within Research Project No. 18-12-00169-p.

\bibliographystyle{apsrev}
\bibliography{intro,calcdetails,lit}

\end{document}